\documentclass[twocolumn]{aastex631}

\hypersetup{linkcolor=red,citecolor=blue,filecolor=cyan,urlcolor=magenta}

\usepackage{CJK}
\usepackage{amsmath} 
\usepackage{bm}      
\usepackage{lmodern}
\usepackage{amssymb}
\usepackage{physics} 
\usepackage{pgfplots}  
\pgfplotsset{compat=1.17} 

\shorttitle{Solar Convection and Rotation}
\shortauthors{Chen Haibin et al.}

\graphicspath{{./}{figures/}}

\begin{document}
\begin{CJK*}{UTF8}{gbsn}
\title{The Interaction between Solar Convection and Rotation}

\author[0000-0002-5500-3634]{Haibin Chen (陈海彬)}

\author[0000-0003-0264-4363]{Rong Wu (吴蓉)}
\correspondingauthor{Rong Wu}
\email{wurong2@mail3.sysu.edu.cn}

\begin{abstract}


The rotational energy of a fluid parcel changes during isotropic expansion or compression. In solar convection, rotation absorbs energy from convection and inhibits it, causing the motion of fluid parcels larger than a critical size to become vibration. Turbulence and inertial oscillations can cause the deformation of fluid parcels to deviate from isotropic, altering the equilibrium position of the vibration and forming motion larger than the critical size, respectively, the large granules within the granules and probably the mesogranulation. 

The change in rotational energy of granules during convection causes their rotation speed to differ from the local speed, forming a statistically significant solar radial differential rotation. The meridional circulation driven by radial differential rotation transports angular momentum towards the equator, forming the latitudinal differential rotation. 

A model constructed by combining mixing length theory explains why granule size and temperature distribution are independent of latitude, and the structure produced by this mechanism is similar to the characteristics of supergranules.


\end{abstract}



%

\section{Introduction}

There are still many unresolved issues regarding granules, mesogranules, supergranules,\citep{2003ApJ...597.1200R} and solar differential \citep{2010Ap&SS.328..269P}.

Granules have been shown to be of convective origin\citep{1930ZA......1..138U} \citep{1950ApJ...111..351R}, and they can be classified into small and large granules   based on a diameter of approximately 1''37\citep{1986SoPh..107...11R}, with significantly different number distributions, fractal dimensions, and brightness distributions between them\citep{1997ApJ...480..406H}. However, there is still no widely accepted explanation for this phenomenon.

The origin of mesogranules \citep{1981ApJ...245L.123N}  is even debated. Mesogranular intensity fluctuations are reported to have correlations with vertical velocity consistent with convection\citep{1989A&A...216..259D}\citep{1992A&A...256..652S}.

Supergranules are primarily composed of horizontal motions\citep{1954MNRAS.114...17H}\citep{1962ApJ...135..474L}, with enhanced spectral intensity at cell boundaries\citep{1968SoPh....5..309B}\citep{1970SoPh...14...89F}, suggesting that they may not be of convective origin.

The equatorial rotation rate of the Sun is faster than that at the poles\citep{1905MNRAS..65..813M}\citep{1951MNRAS.111..413N}\citep{1965Obs....85...37W}\citep{1974ARA&A..12...47G}. Most theoretical explanations for this phenomenon have failed, and simulations are now commonly used to explain it\citep{2010Ap&SS.328..269P}.

In our study of differential rotation, we have found that rotation can inhibit convection\citep{2022arXiv220711990C}. Representing changes in rotational energy as equivalent temperatures can facilitate understanding and research\citep{2022arXiv221108113C}. However, this does not explain the following questions: why large granules exceeding a critical scale can exist and exhibit turbulence-like characteristics, how mesogranules and supergranules form under the inhibition of rotation, why the size  of granules and temperature\citep{1988Sci...242..908K}\citep{1988Natur.332..810W}\citep{1993JGR....9818895H} at the solar poles and equator are so similar, and how solar differential rotation extracts energy from granular convection and forms.

%
%
%
%
%
%
%
%
%

\section{Rotational Equivalent Temperature}

The rotation of a fluid alters its pressure distribution. For typical shapes such as spheres, cylinders, and cubes, rotation generally increases their surface pressure. As an example, consider a fluid cylinder rotating about its axis of symmetry at a speed of $\Omega$, with a radius of $l$ and a height of $2l$, and a fluid density of $\rho$. The average increase in surface pressure due to rotation is given by:

\begin{equation}
{\bar{p}}_\mathrm{\Omega}=\frac{1}{6}\rho\mathrm{\Omega}^2l^2
\end{equation}

The change in pressure distribution caused by rotation raises a question: when the shape of the fluid changes, does the rotational energy also change? For instance, in the most typical isotropic expansion process, density, rotational speed, and rotational pressure all vary with the radius $l$, and there is a correlation between them. The relationship between rotational pressure and density is:

\begin{equation}
\frac{d{\bar{p}}_\mathrm{\Omega}}{{\bar{p}}_\mathrm{\Omega}}=\frac{2}{3}\frac{d\rho}{\rho}
\end{equation}
This relationship is similar to that between gas pressure and density with a degree of freedom of 3. Therefore, the effects of rotation can also be described using an equivalent rotational temperature, which facilitates the inclusion of rotational energy effects in convection criteria. Following the gas state equation, let ${\bar{p}}_\mathrm{\Omega}=\rho\frac{R_m}{M_m}T_\Omega$. The equivalent rotational temperature is then:

\begin{equation}
T_\Omega=\frac{M_m}{6R_m}\Omega^2l^2
\end{equation}
For different shapes and axes of rotation, such as spheres, squares, or cylinders with axes of rotation that are not symmetric, the coefficients of additional pressure and equivalent temperature generated by rotation differ. Let:

\begin{equation}
T_\Omega=k_\Omega\frac{M_m}{R_m}\Omega^2l^2
\end{equation}
Where the value of $k_\Omega$ depends on the shape and axis of rotation, and can be calculated based on changes in rotational energy during isotropic expansion.

In fluid mechanics, the pseudo-vorticity energy $\frac{\omega^2}{2}$ is used to describe the intensity of fluid rotation in a certain region, and $\frac{\Omega^2}{8}$ is equivalent to pseudo-vorticity energy. In the formula, $\Omega=\left|\vec{\Omega}\right|$ is a scalar describing the magnitude of the fluid parcel's rotational speed. In rotating bodies like the Sun, using rotational speed is more intuitive and convenient than using vorticity, and it avoids conceptual pitfalls, allowing us to perform seemingly unreasonable operations without psychological burden but still obtain results that align with observed data.

The rotation of fluid parcels affects their deformation through inertial oscillations, and the magnitude of the response period is the same as the Sun's rotation period $T_0$. There is a difference of 4-5 orders of magnitude between the existence time of granulation and $T_0$, so it can be considered that the expansion and compression of granules have not yet been affected by rotation, and are statistically close to isotropic. The equivalent rotational temperature performs well in this context. However, for mesogranules and supergranules, the effects of inertial oscillations need to be considered.

\section{Convection Criterion in the Solar Convection Zone}  

Under normal conditions, we can use the Schwarzschild convection criterion to determine whether convection will occur spontaneously:
\begin{equation}
\left|\frac{dT}{dR}\right|_{rd}>\left|\frac{dT}{dR}\right|_{ad}
\end{equation}
Where $\left(\frac{dT}{dR}\right)_{rd}$ is the temperature gradient of the fluid along the solar diameter, and $\left(\frac{dT}{dR}\right)_{ad}=\left(\gamma-1\right)T\frac{d\rho}{\rho dR}$ is the adiabatic temperature gradient.

In solar convection, for spherically symmetric expanding (compressing) fluid parcels, when considering the influence of rotational energy considerations, the convection criterion becomes:
\begin{equation}
\left|\frac{dT+dT_\Omega}{dR}\right|_{rd}>\left|\frac{dT+dT_\Omega}{dR}\right|_{ad}
\end{equation}
Similarly, $\ \left(\frac{dT_\Omega}{dR}\right)_{ad}=\frac{2}{3}T_\Omega\frac{d\rho}{\rho dR}$ represents the change in $T_\Omega$ during the spherically symmetric expansion of a fluid parcel that is not affected by viscosity. We note that in this process, when the relationship $\frac{d\Omega}{\Omega d R}=\frac{2d\rho}{3\rho dR}$ is satisfied, the fluid parcel satisfies the relationship $\left(\frac{dT_\Omega}{dR}\right)_{rd}=\left(\frac{dT_\Omega}{dR}\right)_{ad}$, which can be used to describe the adiabatic temperature gradient of the rotational equivalent temperature. $\left(\frac{dT_\Omega}{dR}\right)_{rd}$ is the rotational equivalent temperature gradient during the motion of the fluid parcel, which can be expressed using the rotational speed gradient. Under the same rotational speed gradient, it is proportional to the square of the fluid parcel size $l$, so the convection criterion is affected by the size of the fluid parcel.

In the solar convection zone, generally, the temperature gradient satisfies the convection criterion, i.e., $\left|\frac{dT}{dR}\right|_{rd}>\left|\frac{dT}{dR}\right|_{ad}$, but the rotational speed gradient does not satisfy the convection criterion. In most regions of the solar convection zone, $\left|\frac{dT}{dR}\right|_{rd}<\left|\frac{dT}{dR}\right|_{ad}$. In this case, the size $l$ of the fluid parcel can determine whether its own motion is promoted or inhibited. The critical size $l_{ad}$ satisfies:
\begin{equation}
l_{ad}^2=\frac{-T\left(\frac{dT}{TdR}-\left(\gamma-1\right)\frac{d\rho}{\rho dR}\right)}{\frac{{{2k}_\Omega M}_m}{R_m}\Omega^2\left(\frac{d\Omega}{\Omega d R}-\frac{2}{3}\frac{d\rho}{\rho dR}\right)}
\end{equation}
The convection criterion is:
\begin{equation}
l<l_{ad}
\end{equation}
This convection criterion only holds for spherically symmetric expanding or compressing fluid parcels. When the size of the fluid parcel satisfies the convection criterion, the fluid parcel is in a state of natural convection. When $l>l_{ad}$, the motion of the fluid parcel is inhibited.

From the perspective of energy flow, thermal energy flows from high-density regions to low-density regions, accompanied by energy release. Rotational kinetic energy flows from low-density regions to high-density regions, which is an energy absorption process. Moreover, the larger the fluid parcel, the more energy per unit mass it absorbs. When the size of the fluid parcel is $l=l_{ad}$, the release of thermal energy and the absorption of rotational kinetic energy reach a balance.

The theory of rotational equivalent temperature can solve the formation and distribution of small granules in granules to a certain extent, but it raises a large number of new questions:

1. Structures of the size of large granules within granules, mesogranules and supergranules  should be in a state of oscillation according to this theory. Why can they form motions?

2. A portion of the energy in convection is converted into the rotational kinetic energy of fluid parcels. Are they related to the formation of solar differential rotation?

3. The speed of solar rotation decreases significantly in high-latitude regions. According to this theory, the size of granules should also increase significantly, but observations show that there is no significant difference in granule size along the latitude. What causes this?

The establishment of new theories requires us to immediately address the above issues, so next, we will provide a simple and reasonable explanation for them.

\section{Vibration Dominated by Rotation}

In the solar convection zone, when the size of a spherically symmetric deformed fluid parcel exceeds the critical size $l > l_{ad}$, the equation of motion for the fluid parcel has solutions in the form of vibrations. We hope to obtain the specific equation of motion, which serves as a validation of previous inferences and is necessary for studying the motion of other structures.

Based on the ideal gas law $p = \frac{M_m}{R_m}\rho T$, the equation of state for a rotating fluid parcel can be written as:
\begin{equation}
p + \bar{p}_\Omega = \frac{M_m}{R_m}\rho(T + T_\Omega)
\end{equation}
After the fluid parcel rises and expands, the pressure remains in equilibrium with the surrounding fluid. The difference in density between the fluid parcel and its environment is given by:
\begin{equation}
\mathrm{\Delta}\rho = - \rho\frac{\mathrm{\Delta}T + \mathrm{\Delta}T_{\Omega}}{T + T_{\Omega}}
\end{equation}
The motion of the fluid parcel is driven by the difference between gravity and buoyancy:
\begin{equation}
\frac{d^{2}R}{{dt}^{2}} - \frac{\mathrm{\Delta}\rho}{\rho}g = 0
\end{equation}
Upon rearrangement, we obtain:
\begin{equation}
\frac{d^2R}{{dt}^2} + \frac{T\left(\frac{2}{3}\frac{d\rho}{\rho dR} - \frac{dT}{T dR}\right) + 2T_\Omega\left(\frac{2}{3}\frac{d\rho}{\rho dR} - \frac{d\Omega}{\Omega dR}\right)}{T + T_\Omega}g\delta R = 0
\end{equation}
Where $\delta R = R - R_0$ and $R_0$ is the equilibrium position. Letting $\lambda^2 = -\frac{T\left(\frac{2}{3}\frac{d\rho}{\rho dR} - \frac{dT}{T dR}\right) + 2T_\Omega\left(\frac{2}{3}\frac{d\rho}{\rho dR} - \frac{d\Omega}{\Omega dR}\right)}{T + T_\Omega}g$, the solution to the equation is:
\begin{equation}
\delta R = C_1e^{\lambda t} + C_2e^{-\lambda t}
\end{equation}
Where $T_\Omega$ varies with the size of the fluid parcel $l$ and affects the value of $\lambda$. The value of $\lambda$ can be expressed in terms of $l$ as:
\begin{equation}
\lambda^2 = -\frac{2\frac{T_\Omega}{l^2}\left(\frac{2}{3}\frac{d\rho}{\rho dR} - \frac{d\Omega}{\Omega dR}\right)(l^2 - l_{ad}^2)}{T + T_\Omega}g
\end{equation}
When the size of the fluid parcel $l < l_{ad}$, $\lambda^2 > 0$, and the solution to the equation is an exponential function depending on the initial conditions, similar to the state of thermal convection. When $l > l_{ad}$, the equation has solutions in the form of vibrations. This is a specific manifestation of the convective criterion considering the equivalent rotating temperature.

In particular, when $l \gg l_{ad}$, we have:
\begin{equation}
\lambda \approx \sqrt{-2\left(\frac{2}{3}\frac{d\rho}{\rho dR} - \frac{d\Omega}{\Omega dR}\right)g}
\end{equation}
This can be used to estimate the vibration period $t_{vib}$ associated with structures of scales such as mesogranules and supergranules.

Among these, we can let $\frac{d\Omega}{\Omega dR} \approx 0$ and $g_{s} = - 274.m/s^{2}$. The density distribution of the solar convection zone can be simply approximated as $\rho \approx C_\rho\left(\frac{R_\odot}{R} - 1\right)^\frac{1}{\gamma - 1}$, where $C_\rho = 0.6g/{cm}^{3}$ and $\gamma = \frac{5}{3}$. The results obtained agree with more accurate solar models within $\pm10\%$. We find that $\frac{d\rho}{\rho dR} = -\frac{3}{2}\frac{R_\odot}{\left(\frac{R_\odot}{R} - 1\right)R^2} \approx -\frac{3}{2}\frac{1}{\left(R_\odot - R\right)}$. If the shallowest depths of mesogranules and supergranules are the same as their radii, i.e., $\left(R_\odot - R\right)$ is taken as $3Mm$ and $16Mm$ respectively\citep{2003ApJ...597.1200R}, then the minimum vibration period of mesogranules is about $7.7min$, and that of supergranules is about $18min$. Their typical vibration periods are shorter than their lifetimes, suggesting that their motion may be a shift in equilibrium position. The vibration period of large granules is significantly affected by $l$ and is close to $5min$. This period is similar to the observed $5min$ solar oscillation\citep{1962ApJ...135..474L}\citep{1989nos..book.....U}\citep{1995SoPh..162..129S}, indicating that rotation-dominated vibrations may be an important entry point for helioseismic research.

If it is assumed that there is an exchange of thermal energy between the fluid parcel and the external environment, the equation of motion for a fluid parcel with a size $l > l_{ad}$ has a solution with excited vibrations. Large granules are the smallest moving structures with vibrational patterns, and thermal exchange occurs rapidly, so excited vibrations are more pronounced in large granules.

%
%
%
%
%
%
%
%

\section{lage granules and the equilibrium position jump generated by turbulence}

We now investigate the motion and characteristics of large granules within granules.

The stretching of fluid parcels by turbulence leads to variations in the rotation rate $\Omega$, altering the $T_\Omega$ of the fluid parcel. The equilibrium position jumps to anther place , and the distance of this jump is primarily determined by the change in $\Omega$ and is also influenced by temperature gradients. This will cause the fluid parcel, which is initially at rest, to move. This change is reflected in the equation of motion $\delta R = C_1e^{\lambda t} + C_2e^{-\lambda t}$, where significant modifications occur to the coefficients $C_1$ and $C_2$ even with small variations in $\lambda$.

Fluid parcels with size $l < l_{ad}$ are in an unstable equilibrium state. The jump in equilibrium position is equivalent to a relatively strong disturbance to the thermal convection, promoting its formation.

Fluid parcels with size $l > l_{ad}$ are in a stable equilibrium state. After the jump in equilibrium position, the originally stationary fluid parcel begins to oscillate around the new equilibrium position, forming a large granule. The distance of the jump in equilibrium position is amplified by temperature gradients, and the closer the size is to $l_{ad}$, the greater the amplification. Large granules can form excited oscillations with the involvement of thermal transport.

New thermal convection can also be generated within a large granule, which may roughen the boundaries of fluid parcels. Thermal convection and oscillation can even be relatively independent, and in some cases, changes in the direction of motion during oscillation may be interpreted as the fragmentation of large granules. This explains why the fractal dimension of large granules is significantly greater than that of small granules. Due to its oscillatory motion, it exhibits notably different characteristics compared to thermal convection. For example, its brightness is primarily related to the amplitude of oscillation and not to its size.

The jump distance of the equilibrium position of large granules affects the amplitude of oscillation. It is strongly influenced by the turbulence energy spectrum, and the amplitude of larger granules is almost entirely controlled by the turbulence energy spectrum. This may make its energy spectrum closely resemble that of turbulence.

\section{The inertial oscillation and the motion generated by the shift of the equilibrium position of the fluid parcels}

When studying solar granulations, we consider their expansion and compression to be statistically close to spherically symmetric. However, when the magnitude of the existence time of fluid parcels approaches the solar rotation period, such as in mesogranulations and supergranulations, the imbalance of pressure within the fluid parcels can cause anisotropic deformation, known as inertial oscillation. At this point, the deformation of the fluid parcels is no longer statistically isotropic, and the equilibrium position of their vibration is influenced by non-spherical symmetric deformation. Since the vibration period of the fluid parcels is much smaller than the period of inertial oscillation, the movement speed of the equilibrium position $R_0$ can represent the average motion speed of the fluid parcels.

According to the characteristics of inertial oscillation, the period $T$ of inertial oscillation of fluid parcels should be close to half of the period $T_0$ of the rotating system. During the process of inertial oscillation of fluid parcels, assuming uniform internal rotation rate denoted as $\Omega$, and the environmental fluid rotation rate as $\Omega_0$, the natural angular frequency of undamped vibration is $\omega_n=2\Omega_0$. Based on the conservation of angular momentum and the change in radius perpendicular to the axis of rotation of the fluid parcels, the equation for the change in rotation rate of the fluid parcels can be derived as:
\begin{equation}
\frac{\partial^2 \Omega}{\partial t^2} +2\zeta\omega_n \frac{\partial \Omega}{\partial t}+\omega_n^2 (\Omega-\Omega_0 )=0
\end{equation}
where $\zeta$ is the damping ratio, mainly caused by the propagation of inertial waves. If $\frac{\partial \Omega}{\partial t}$ is small, this term can be neglected.

According to the equation of motion for fluid parcels, at any time $t$ during the motion of fluid parcels, the rotation rate of the fluid parcels is $\Omega$, the environmental fluid rotation rate is $\Omega_0$, and the fluid parcels are in a state of forced vibration equilibrium at position $R_0$ with $\delta\rho=0$. This leads to the equation:
\begin{equation}
\Delta T_{\Omega}+\Delta T=0
\end{equation}
This implies that there exists a rotational speed difference between the fluid parcels and the environment determined by $\Delta T$:
\begin{equation}
\mathrm{\Delta}\Omega = \Omega - \Omega_{0} = - \frac{\mathrm{\Delta}T}{{2k}_{\Omega}\frac{M_{m}}{R_{m}}\Omega l^{2}}
\end{equation}
The rotational speed difference gives rise to inertial oscillations in the fluid parcels, resulting in anisotropic deformation and changes in rotational speed. Denoting the change in rotational speed caused by inertial oscillations as $\frac{\partial \Omega}{\partial t}$, initially, $\frac{\partial \Omega}{\partial t}$ is small and varies little with density. Neglecting damping effects, the equation for rotational speed oscillations of the fluid parcels can be simplified to:
\begin{equation}
\frac{\partial^2 \Omega}{\partial t^2} +\omega_n^2 \Delta \Omega=0
\end{equation}
Substituting the equation for rotational speed difference and integrating, we obtain:
\begin{equation}
\frac{\partial\Omega}{\partial t} = {\int_{0}^{t}{\frac{\omega_{n}^{2}\mathrm{\Delta}T}{{2k}_{\Omega}\frac{M_{m}}{R_{m}}\Omega l^{2}}dt}}
\end{equation}
The change in rotational speed caused by inertial oscillations, $\frac{\partial \Omega}{\partial t}$, leads to a change in the equilibrium position of forced vibration, $R_0$. At time $t+dt$, the fluid parcels reach a new equilibrium position $R_0^\ast$. Within the time $dt$, both density changes and inertial oscillations within the fluid parcels affect their rotational speed, and rotational speed gradients affect the rotational speed of the environmental fluid. At the new equilibrium position $R_0^\ast$, the rotational speed difference between the fluid parcels and the environmental fluid is given by:
\begin{equation}
\begin{split}  
\Delta \Omega^\ast &=\Omega^\ast-\Omega_0^\ast \\
&=(\Omega+\frac{\partial \Omega}{\partial t} dt+\frac{2}{3} \Omega \frac{\partial \rho}{\rho \partial R}  \frac{\partial R_0}{\partial t} dt)-(\Omega_0+\frac{\partial \Omega}{\partial R}  \frac{\partial R_0}{\partial t} dt)
\end{split}  
\end{equation}
where the term $\frac{\partial \Omega}{\partial t} dt$ represents the change in rotational speed caused by anisotropic deformation of the fluid parcels, related to rotational speed oscillations; the term $\frac{2}{3} \Omega \frac{\partial \rho}{\partial R}  \frac{\partial R_0}{\partial t} dt$ represents the change in rotational speed caused by isotropic deformation of the fluid parcels, related to density changes; and the term $\frac{\partial \Omega}{\partial R}  \frac{\partial R_0}{\partial t} dt$ represents the change in rotational speed of the environmental fluid, related to rotational speed gradients. The rotational speed difference $\Delta \Omega^\ast$ between the fluid parcels and the environment is determined by $\Delta T$.

Combining the above equations, we obtain:
\begin{equation}
\frac{\partial R_{0}}{\partial t} = \frac{\frac{\partial\Omega}{\partial t}}{\frac{\partial\Omega}{\partial R} - \frac{2}{3}\Omega\frac{\partial\rho}{\rho\partial R}} = \frac{\int_{0}^{t}{\frac{\omega_{n}^{2}\delta T}{{2k}_{\Omega}\frac{M_{m}}{R_{m}}\Omega l^{2}}dt}}{\frac{\partial\Omega}{\partial R} - \frac{2}{3}\Omega\frac{\partial\rho}{\rho\partial R}} = K{\int_{0}^{t}{\frac{\delta T}{l^{2}}dt}}
\end{equation}
Where,
$
K = \frac{\omega_{n}^{2}}{{2k}_{\Omega}\frac{M_{m}}{R_{m}}\Omega^{2}\left( {\frac{\partial\Omega}{\Omega\partial R} - \frac{2}{3}\frac{\partial\rho}{\rho\partial R}} \right)}
$
Taking the derivative of the above equation, we get:
\begin{equation}
\frac{\partial^2 R_0}{\partial t^2} = K \frac{\delta T}{l^2}
\end{equation}
This is the equation of motion for a convective structure with a size significantly larger than the critical scale during its initiation phase. The coefficient $K$ depends on the system's parameters. It should be noted that this equation does not include the term for gravitational acceleration because the focus of this problem is on the rotational oscillation of fluid parcels. Gravity only assists in creating a rotational difference between the fluid parcels and their environment through the temperature difference $\delta T$ and helps the fluid parcels reach a new equilibrium position in the subsequent process, maintaining the existence of the rotational difference.

Since the oscillation periods of mesogranules and supergranules are generally much shorter than their lifetimes and inertia oscillation periods, the equation of equilibrium position change may represent their equation of motion.

\section{Size of Moving Fluid Parcels Generated by Inertial Oscillations}

The diameters of supergranules concentrate around 32Mm, while those of mesogranules concentrate around 7Mm. The motion induced by inertial oscillations may be one of them, and we can attempt to analyze the reasons for their concentrated size distribution.

In the equation of motion for mesogranules or supergranules, the environmental parameter $K$ depends only on the environment and is a given parameter. By supplementing the equation with the temperature difference $\delta T$ between the fluid parcel and the environment and the size $l$ of the fluid parcel, their motion can be solved.

If the fluid parcel does not disintegrate or add matter during its motion, then the mass of the fluid parcel is conserved, and the size $l$ of the fluid parcel satisfies the relationship:
\begin{equation}
l^3 \rho = l_0^3 \rho_0
\end{equation}

The temperature difference $\delta T$ between the fluid parcel and the environment is related to heat conduction. Under conditions of sufficient heat conduction or adiabatic conditions, $\delta T$ can be easily solved.

Let $\kappa$ be the heat conduction coefficient including heat conduction and turbulent heat transfer. When $l \gg 4\sqrt{\kappa t}$, the heat transport process can only affect the area near the boundary of the fluid parcel within a range of $4\sqrt{\kappa t}$, and the fluid parcel can be considered adiabatic. Let $R_{00}$ be the initial equilibrium position, and $\delta T$ satisfies the relationship:
\begin{equation}
\delta T = \left( \frac{\partial T}{\partial R} - (\gamma - 1) T \frac{\partial \rho}{\rho \partial R} \right) (R_{00} - R_0)
\end{equation}
So when $l \gg 4\sqrt{\kappa t}$, the equation of motion for the equilibrium position is:
\begin{equation}
\frac{\partial^2 R_0}{\partial t^2} = \frac{K \left( \frac{\partial T}{\partial R} - (\gamma - 1) T \frac{\partial \rho}{\rho \partial R} \right) (R_{00} - R_0)}{l_0^2 \left( \frac{\rho_0}{\rho} \right)^{2/3}}
\end{equation}
In this equation, apart from $l_0^2$ and the initial position $R_{00}$ of the fluid parcel, the other parameters are all environmental. For fluid parcels generated simultaneously in the same region, the smaller their size $l$, the greater their acceleration. After a period of time, the displacement and $\delta T$ become relatively large, which inhibits the motion of larger fluid parcels. Therefore, when $l \gg 4\sqrt{\kappa t}$, fluid parcels tend to have smaller sizes $l$.

When $4\sqrt{\kappa t_{\text{vib}}} \ll l \ll 4\sqrt{\kappa t}$, the heat transport during the equilibrium position movement is sufficient, and the heat conduction process has not yet destroyed the vibration process. In this case, $\delta T$ satisfies:
\begin{equation}
\delta T = K_2 l^2 v
\end{equation}
where $v$ is the velocity of the fluid parcel's equilibrium position movement, and $K_2$ is determined by the shape of the fluid parcel and environmental parameters. The equation of motion for supergranules is:
\begin{equation}
\frac{\partial^2 R_0}{\partial t^2} = K K_2 \frac{\partial R_0}{\partial t}
\end{equation}
It can be seen that the velocity of the fluid parcel's motion is independent of its size. However, for fluid parcels generated simultaneously in the same region, the larger $l$ is, the greater $\delta T$ becomes, which inhibits the motion of relatively smaller fluid parcels. Therefore, when $l \ll 4\sqrt{\kappa t}$, fluid parcels tend to have larger sizes $l$.

In summary, the radii of fluid parcels tend to concentrate around $4\sqrt{\kappa t}$.

Observations show that in addition to granulation, there are two convective structures in the sun that are significantly larger than granules: mesogranules and supergranules. The above model can only explain the formation of one of these structures. Supergranules mainly move horizontally, while mesogranules have significant convective characteristics. Therefore, we suggest that mesogranules may be caused by the movement of oscillating equilibrium positions.

\section{Mechanism of Radial Differential Rotation}

The rotation traps a portion of the energy from thermal convection and suppresses it, which causes the rotation rate of fluid parcels to differ from the local rotation rate. A large number of fluid parcels with different rotation rates can form a statistical difference in rotation rates, known as differential rotation.

To analyze this more specifically: when a fluid parcel moves outward along the solar diameter, it expands and its rotation rate decreases. When it moves inward along the solar diameter, it compresses and its rotation rate increases. Therefore, the rotation rate of fluid parcels at the top of the convection zone is lower than that at the bottom.

Next, let's analyze the radial distribution of fluid parcel rotation rates. Ignoring the effects of viscosity on fluid parcel rotation, and assuming that the existence time of fluid parcels is significantly less than the solar rotation period and that the fluid parcels have not disintegrated, the rotation rate of fluid parcels is only related to the initial rotation rate $\Omega_0$, initial density $\rho_0$, and current density $\rho$. That is,
\begin{equation}
\Omega=\Omega_0 (\rho/\rho_0 )^{2/3}
\end{equation}
After sufficient mixing, the average rotation rate of fluid parcels in a region is related to the average density and rotation rate of the source region of the fluid parcels, which is given by
\begin{equation}
\Omega=(\sum_{i=1}^n\Omega_i (\rho/\rho_i )^{2/3} )/n
\end{equation}
In the edge region of the convection zone, fluid parcels mainly come from inside the convection zone, and there are significant differences in the average density between the source region and the local region. For example, fluid parcels at the top of the convection zone come from regions with higher densities inside the convection zone, so the rotation rate of fluid parcels at the top of the convection zone is lower, and correspondingly, the rotation rate of fluid parcels at the bottom of the convection zone is higher. If we know the radial distribution of the free path of fluid parcels, we can calculate the above formula.

The different radial rotation rates of fluid parcels can form a statistical velocity difference. The rotation of fluid parcels and their irregular translation are relatively independent. In cylindrical coordinates $(r,\theta,z)$, let the statistical velocities along the $\theta$ direction generated by the translation and rotation of fluid parcels be $v_{\theta ta}$ and $v_{\theta ro}$, respectively. The statistical velocity along the $\theta$ direction $v_\theta$ is given by
\begin{equation}
v_\theta=v_{\theta ta}+v_{\theta ro}
\end{equation}
In general, if irregular motion dominates, we have $v_{\theta ta}=\Omega_0 r$, where $\Omega_0$ is a fixed rotation rate. If the rotation rates of all fluid parcels are the same, then $v_{\theta ro}=0$. Combining these, we can obtain a solution without differential rotation, $v_\theta=\Omega_0 r$. Most previous models attributed differential rotation to $v_{\theta ta}$ because the irregular motion of fluid parcels in turbulence can be destroyed by the dynamics of thermal convection, but the calculated results did not match observations. In our model, we believe that radial differential rotation originates from differences in the rotation of fluid parcels, $v_{\theta ro}$, and we can ignore the contribution of $v_{\theta ta}$ to differential rotation. We define their difference as $\delta v_\theta=v_\theta-\Omega_0 r\approx v_{\theta ro}$.

When the rotation rates of fluid parcels in different regions are not the same, $v_{\theta ro}$ is no longer zero. To simplify the calculation and make the results more intuitive, we approximate each fluid parcel as six particles located on both sides of three mutually perpendicular lines passing through the center of the fluid parcel, with a distance of a from the center. During statistical analysis, these three lines are parallel to the coordinate axes of the cylindrical coordinate system $(r,\theta,z)$. The outer particles are located at a radius of $(r+l)$, while the inner particles are located at $(r-l)$. The rotation rate of the fluid parcel affects the $v_{\theta ro}$ of these two points. Correspondingly, the $v_{\theta ro}$ of the fluid at position r is influenced by the rotation rates of fluid parcels at positions $(r+l)$ and $(r-l)$. The velocity of the fluid parcel at $(r-l)$ falling onto the point at r is given by
\begin{equation}
v_{\theta ro-}=l\Omega_{(r-l)}
\end{equation}
where $\Omega_{(r-l)}$ represents the rotation rate of the fluid parcel located at $(r-l)$. Similarly, we can obtain the $v_{\theta ro}$ of other points, which is given by

\begin{equation}
\begin{split}  
v_{\theta ro}&=\frac{1}{6} (v_{\theta r+}+v_{\theta r-}+4\delta v_{\theta 0} ) \\
&=\frac{1}{6} (l\Omega_{(r-l)}-l\Omega_{(r+l)} )
\approx-\frac{1}{3} l^2  \frac{\partial\Omega}{\partial r}
\end{split}  
\end{equation}
It should be noted that $\Omega$ here refers to the average rotation rate of fluid parcels, which may differ from the average rotation rate of the environment. When the gradient of the average rotation rate of fluid parcels, $\frac{\partial\Omega}{\partial r}$, is not zero, the value of $v_{\theta ro}$ is not zero, resulting in differential rotation. This is the mechanism by which radial differential rotation occurs in the sun.

\section{A Simple Model of Radial Differential Rotation}

Previously, only the calculation method for the distribution of fluid parcel rotation rates was given, but due to the complexity of the actual calculations, no specific distribution was provided. The true distribution of fluid parcel rotation rates can be obtained through numerical simulations. However, by using a greatly simplified model, we can obtain an intuitive and specific distribution of radial differential rotation that differs from the real results but can assist in studying the formation of radial differential rotation on the solar equator.

We construct a thin annular cylinder model with the z-axis as the axis of symmetry. The radius of the fluid parcel is $l$, the thickness of the annular cylinder is $2b$, the inner radius of the annular cylinder is $r_1-b$, and the outer radius is $r_1+b$. There is no restriction in the z-direction, and the inside and outside of the annular cylinder are filled with compressible fluid. The boundary of the annular cylinder is set as the convective boundary. The fluid parcels within the annular cylinder have convection, manifesting as irregular motion, and their centers of mass rebound at the boundary. The fluid parcels outside the annular cylinder have no motion. Assuming a density gradient of $\frac{\partial \rho}{\partial r}$ and to keep the model within a linear range, we set $b \ll r_1$ and $\frac{\partial \rho}{\rho \partial r} b \ll 1$.

If the fluid parcels all originate from the same source, the distribution of their rotational speeds along the radial direction of the cylinder simplifies to:
\begin{equation}
\Omega(r) = \Omega_1 \left( \frac{\rho(r)}{\rho_1} \right)^{\frac{2}{3}}
\end{equation}
This requires that the fluid parcels in any region have not undergone substantial disintegration before they are uniformly distributed throughout the convection zone, and this process takes significantly less time than the solar rotation period. In this equation, $\Omega_1$ and $\rho_1$ represent the average rotation rate and density of the convection zone, respectively.

To further simplify the model, we assume that the change in $\rho(r)$ is small and that the density distribution is nearly linear. We define $\delta \Omega(r) = \Omega(r) - \Omega_1$ and $\delta \rho(r) = \rho(r) - \rho_1 = \frac{\partial \rho}{\rho_1 \partial r} (r - r_1)$ to obtain:
\begin{equation}
\delta \Omega(x) =
\begin{cases}
0 & r < r_1 - b \\
\frac{2}{3} \Omega_1 \frac{\partial \rho}{\rho_1 \partial r} (r - r_1) & r_1 - b < r < r_1 + b \\
0 & r > r_1 + b
\end{cases}
\end{equation}
As you can see, this is a relatively simple distribution.

Substituting the relationship for differential rotation and letting $v_0 = -\frac{2}{9} \Omega_1 \frac{d \rho}{\rho dr} l^2$ (where $v_0 > 0$ in the Sun), we get:

\begin{equation}
v_{\theta ro} \approx
\begin{cases}
0 & (r < r_1 - b - l) \\
-\frac{b - l}{2l} v_0 & (r_1 - b - l < r < r_1 - b + l) \\
v_0 & (r_1 - b + l < r < r_1 + b - l) \\
-\frac{b - l}{2l} v_0 & (r_1 + b - l < r < r_1 + b + l) \\
0 & (r > r_1 + b + l)
\end{cases}
\end{equation}

\begin{figure}[htbp]  
	\centering  
	\includegraphics[width=0.5\textwidth, trim=120 400 120 100, clip]{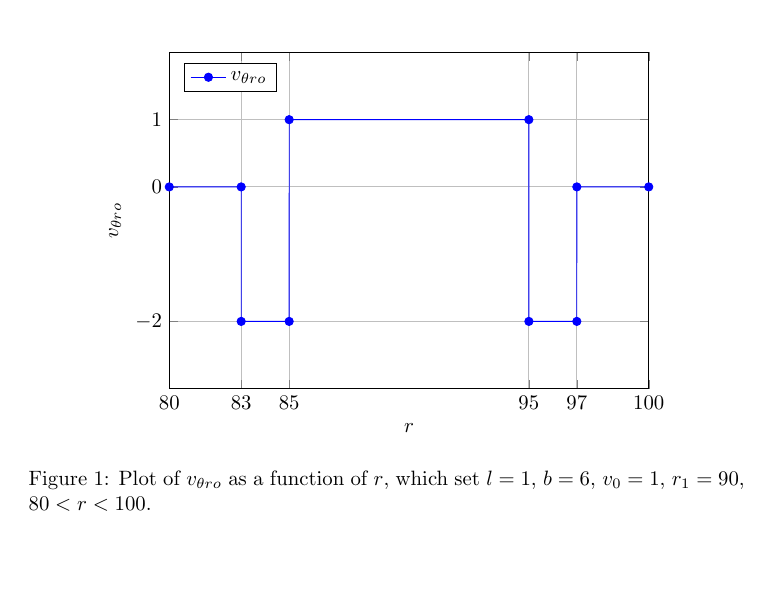}  
	\caption{radial differential rotation}  
	\label{fig:radial}  
\end{figure}  

Statistically, $v_{\theta ro}$ exhibits a strong velocity difference at the convective boundary, with a thickness approximately equal to the diameter of a fluid parcel.\ref{fig:radial}

Outside the convective boundary, since there is no convective transfer of momentum, the distribution of $v_\theta$ is arbitrary. Therefore, the velocity difference at the boundary radii $(r_1 - b - l)$ and $(r_1 + b + l)$ can be arbitrary. Except for the boundary, $\delta v_\theta = v_\theta - \Omega_0 r \approx v_{\theta ro}$.

\begin{equation}
\delta v_\theta \approx
\begin{cases}
v_1 & (r < r_1 - b - l) \\
-\frac{b - l}{2l} v_0 & (r_1 - b - l < r < r_1 - b + l) \\
v_0 & (r_1 - b + l < r < r_1 + b - l) \\
-\frac{b - l}{2l} v_0 & (r_1 + b - l < r < r_1 + b + l) \\
v_1 & (r > r_1 + b + l)
\end{cases}
\end{equation}

This velocity distribution can handle the velocities between the convection zone and the rigid rotation zone at different latitudes, especially in the high-latitude region of the differential rotation layer. It can be seen that the velocity distribution of radial differential rotation generally exhibits an n-shape within the convection zone.

Further refining the model can bring the results closer to observations. For example, considering the free path of fluid parcel disintegration can set the intermediate region of the radial rotation distribution to $v_{\theta ro} = 0$, resulting in a slightly concave feature and an M-shaped velocity distribution. Refining the boundary model to make it a region of rapidly decaying thermal convection rather than a collision boundary can smooth out the distribution of radial differential rotation instead of having sharp transitions.

In regions where the latitude $\varphi$ is not zero, there is an angle between the rotation direction of the fluid parcel and the radial direction of the Sun. After projecting onto the direction of the vertical diameter, the distribution of radial differential rotation can be approximately expressed as:
\begin{equation}
\delta v_\theta (R, \varphi) \approx \frac{\bar{\Omega}(\varphi)}{\bar{\Omega}(0)} \delta v_\theta (R, 0) \cos \varphi
\end{equation}
Where $\bar{\Omega}(\varphi)$ represents the average rotation rate of the convection zone at latitude $\varphi$ and can also be used to approximate the rotational speed of the convection zone around the Sun's axis of rotation.

The radial differential rotation of the Sun is primarily composed of the rotation of fluid parcels, but after the fluid parcels disintegrate, they can retain the characteristics of radial differential rotation. Therefore, a portion of the radial differential rotation is also attributed to the irregular motion of the disintegrated fluid parcels.

\section{latitudinal differential rotation}

The latitudinal differential rotation is a more significant phenomenon than the radial differential rotation. When we project the radial differential rotation onto the axial direction, we found that the resulting axial differential rotation is not zero, which causes an imbalance in the Coriolis force, driving the meridional circulation. The meridional circulation driven by the Coriolis force transports angular momentum away from the axis of rotation (cited from the first paper), making the angular velocity of the fluid outside higher and the inner. This difference in rotational speed is projected onto the latitude, which is the sun's latitudinal differential rotation.

Based on the simple model obtained from the radial differential rotation, a calculable simple model of latitudinal differential rotation can also be further obtained.

The radial differential rotation model is a thin cylinder, which simulates the radial differential rotation on the sun's equatorial plane. To study the latitudinal differential rotation, the model needs to be changed to a thin spherical shell model. If it is assumed that the thickness of the spherical shell is much smaller than the radius of the sphere, the difference in the distance from the convection zone at the same latitude to the axis of rotation can be ignored, and there is no need to repeatedly project in the spherical coordinate system and the cylindrical coordinate system.

The magnitude of the Coriolis force caused by radial differential rotation is
\begin{equation}
a_c=-2\bar{\Omega}\left(\varphi\right)\delta v_\theta\left(R,\varphi\right)
\end{equation}
The rotational speed $\delta v_\theta$ of the upper and lower boundaries of the spherical shell around the sun's axis of rotation is lower than the average, while the rotational speed $\delta v_\theta$ of the middle region of the spherical shell is higher than the average, causing an imbalance in the Coriolis force. This imbalance leads to fluid in the upper and lower boundary regions moving towards high latitudes, and fluid in the middle region moving towards low latitudes, forming meridional circulation. Finally, the latitudinal Coriolis force balances with the viscosity of the meridional circulation. Meridional circulation can affect radial differential rotation, but in this simplified model, we ignore it temporarily.

Assuming that the upper and lower boundaries are free boundaries, and the radial fluid viscosity (including molecular viscosity, turbulent viscosity, and thermal convection) inside the thin spherical shell is $\mu_R$, the equilibrium equation of meridional circulation is
\begin{equation}
\mu_R\frac{\partial^2v_\varphi}{\partial R^2}+\rho a_c\sin\varphi=0
\end{equation}
where $v_\varphi$ is the latitudinal velocity, with the positive direction from low latitude to high latitude.

Letting $\varepsilon={2\frac{\rho}{\mu_R}\frac{\left(\bar{\mathrm{\Omega}}\left(\varphi\right)\right)^2}{\bar{\mathrm{\Omega}}\left(0\right)}v}_0\cos\varphi \sin\varphi$, the velocity distribution $v_\varphi$ is required to be symmetric about $r_1-b$ and continuous. The solution for $v_\varphi$ is
\begin{equation}
v_\varphi\left(\varphi\right)\approx\left\{\begin{matrix}
&-\frac{b-l}{2l}\varepsilon\left(r-r_1+b+l\right)^2+2\frac{b-l}{l}\varepsilon l^2+C_7 \\
&\left(r_1-b-l<r<r_1-b+l\right)\\
\\
&\varepsilon\left(r-r_1\right)^2-\varepsilon\left(b-l\right)^2+C_7 \\
&\left(r_1-b+l<r<r_1+b-l\right)\\
\\
&-\frac{b-l}{2l}\varepsilon\left(r-r_1-b-l\right)^2+2\frac{b-l}{l}\varepsilon l^2+C_7 \\
&\left(r_1+b-l<r<r_1+b+l\right)\\
\end{matrix}\right.
\end{equation}
where $C_7$ can be calculated through mass conservation.

The angular momentum transported along the latitude in time $dt$ is:
\begin{equation}
\begin{split}  
dL&=rdt\iint_{r_1-b-l}^{r_1+b+l}{2\pi r\rho v_\varphi\delta v_\theta d r}\\
&\approx2\pi r^2\rho dt\iint_{r_1-b-l}^{r_1+b+l}{v_\varphi\delta v_\theta d r}\\
&=-2\pi r^2\rho\varepsilon\frac{\bar{\Omega}\left(\varphi\right)}{\bar{\Omega}\left(0\right)}\cos\varphi v_0\left(\frac{8}{3}\left(\frac{b-l}{l}\right)^2l^3+\frac{4}{3}\left(b-l\right)^3\right)dt
\end{split}  
\end{equation}
The angular momentum transported along the latitude generates latitudinal differential rotation, and the latitudinal differential rotation generates angular momentum backflow under the action of viscosity until the total latitudinal angular momentum transport is zero. Letting the latitudinal viscosity coefficient (including molecular viscosity and turbulent viscosity) be $\mu_\varphi$, we have

\begin{equation}
\begin{split}  
&\mu_\varphi\frac{\partial\bar{\mathrm{\Omega}}\left(\varphi\right)}{\partial\varphi}=\frac{dL}{2\pi r\left(2b+2l\right)dt}\\
&=-\frac{r\rho\varepsilon}{\left(2b+2l\right)}\frac{\bar{\Omega}\left(\varphi\right)}{\bar{\Omega}\left(0\right)}\cos\left(\varphi\right)v_0\left(\frac{8}{3}\left(\frac{b-l}{l}\right)^2l^3+\frac{4}{3}\left(b-l\right)^3\right)
\end{split}  
\end{equation}
In the formula, the angular velocity of the fluid around the rotation axis of the Sun and the average angular velocity of the local fluid rotation are treated as a single physical quantity, which may lead to some errors, but they are acceptable. Simplifying, we get
\begin{equation}
\frac{\partial\bar{\mathrm{\Omega}}\left(\varphi\right)}{\partial\varphi}=-\varepsilon_1\left(\bar{\mathrm{\Omega}}\left(\varphi\right)\right)^3\cos^2\varphi \sin\varphi
\end{equation}
where $\varepsilon_\varphi=\frac{r\rho}{\left(2b+2l\right)}\frac{2\rho}{\mu_R}\frac{1}{\bar{\mathrm{\Omega}}\left(0\right)}v_0^2\left(\frac{8}{3}\left(\frac{b-l}{l}\right)^2l^3+\frac{4}{3}\left(b-l\right)^3\right)$. Obviously, $\varepsilon_\varphi>0$.

\begin{figure}[htbp]  
	\centering  
	\includegraphics[width=0.5\textwidth, trim=0 0 0 5, clip]{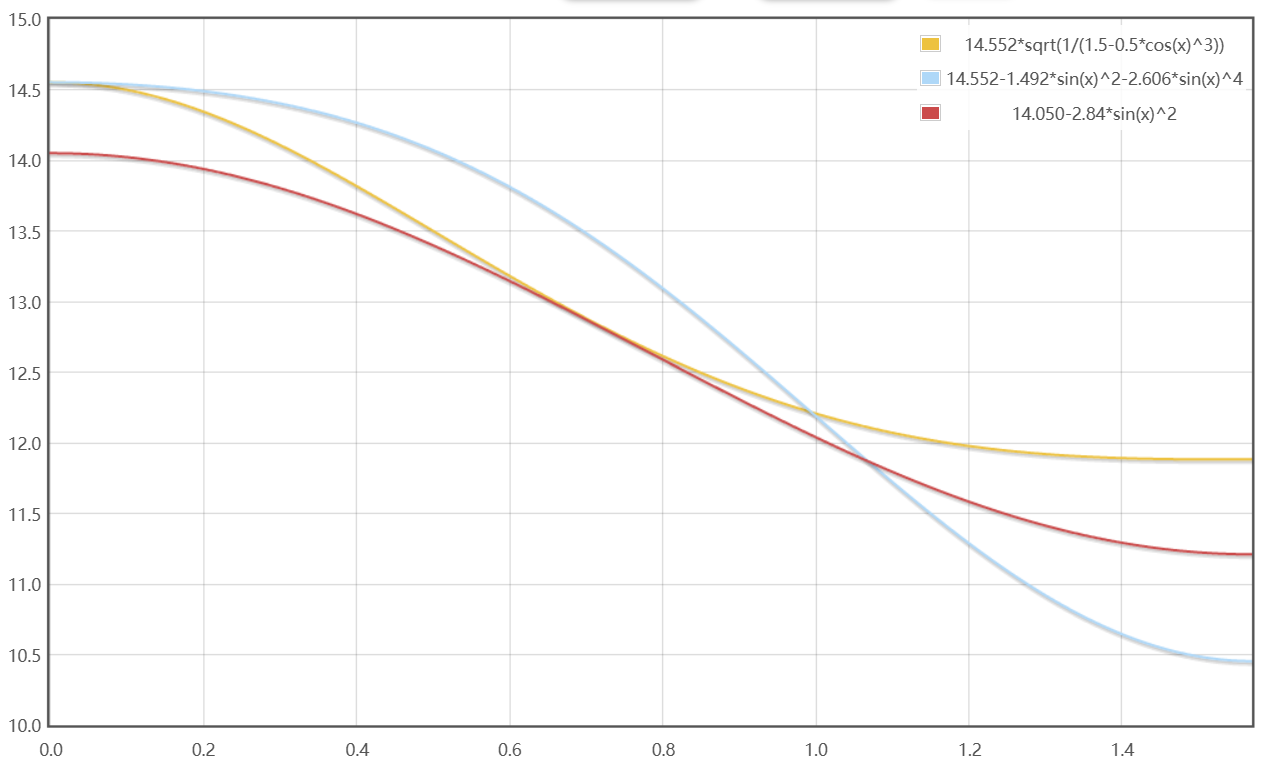}  
	\caption{latitudinal differential rotation}  
	\label{fig:latitudinal}  
\end{figure}

The solution is
\begin{equation}
\bar{\mathrm{\Omega}}\left(\varphi\right)=\sqrt{1/\left(C_\varphi-\varepsilon_\varphi\cos^3\varphi\right)}
\end{equation}
For appropriate values of $C_\varphi$ and $\varepsilon_\varphi$, a comparison with the observed differential rotation on the sun's surface is shown in the figure\ref{fig:latitudinal}.\citep{1984ApJ...283..373H}\citep{1984SoPh...94...13S}

The distribution of latitudinal differential rotation obtained by this formula has the same trend as the observed differential rotation distribution, but there are differences in the specific distribution. This may be due to the simplicity of the model. A more accurate model should be based on a more accurate radial differential rotation model, considering the influence of the thickness of the convection layer and the reaction of meridional circulation on the radial differential rotation.

The Coriolis force experienced by the rotating fluid mass may be entirely borne by the inertial oscillation of the fluid mass itself, and may not drive the differential rotation. Only the fragments of the fluid mass after disintegration can constitute the differential rotation, which can drive the meridional circulation. At this time, the net Coriolis force generated by the radial differential rotation and its driven meridional circulation need to be multiplied by a coefficient much less than 1.

\section{Mixing length theory and convective layer model}

According to the new convection criterion, we can also use other physical quantities to calculate the radial rotation speed gradient at the top of the convection zone. Observations show that the size and temperature of granules do not vary much with latitude, so they can be treated as constants. Therefore, the relative values of the radial rotation speed gradients at different latitudes at the top of the convection zone satisfy the relationship:
\begin{equation}
\Omega^2 \left(\frac{\partial\Omega}{\Omega\partial R} - \frac{2}{3} \frac{\partial\rho}{\rho\partial R}\right) \approx \text{constant}
\end{equation}
Observational data shows that the rotation speed $\Omega$ decreases with increasing latitude, while the radial differential rotation at the top of the convection zone indeed increases with increasing latitude. The observational results are qualitatively consistent with the theory.

This result suggests that the improvement of the new theory requires the reconstruction of the solar convection zone model. After introducing the mixing length theory and assuming that the mixing length is equal to the size of the fluid parcel $l$, there is a closed feedback chain regarding the temperature distribution and granule size. The feedback chain is shown in Figure.

In this feedback chain, the relationship between the rotation speed $\Omega$ and the granule size $l$ (1) is provided by the previous context, the relationship between the granule size $l$ and the temperature distribution $T$ (2) is given by the mixing length theory, the relationship between the temperature distribution $T$ and the linear velocity distribution $v_\theta$ (3) is controlled by the micro-meridional circulation and the Coriolis force, and the relationship between the velocity distribution $v_\theta$ and the rotation speed gradient $\frac{\partial\Omega}{\Omega\partial R}$ (4) is a simple differential. However, it is the relationship (4) that has a strong amplification effect, which mainly comes from the fact that the higher-order differential of small-scale structures is greater than the lower-order, which leads to the maintenance of constant granule size and temperature distribution under significant changes in $\Omega$ at different latitudes.

The specific equations for each relationship are given below.

According to the mixing length theory, when the convective energy flux remains constant, the relationship (2) between the granule size $l$ and the temperature gradient is:
\begin{equation}
\left(\frac{\partial T}{T\partial R} - (\gamma - 1) \frac{\partial\rho}{\rho\partial R}\right)^{\frac{3}{2}} l \approx \text{constant}
\end{equation}
It can be seen that as the granule size increases, the temperature gradient approaches the adiabatic gradient when the energy flux remains constant.

The radial temperature difference between different latitudes can drive micro-meridional circulation, which can change the radial velocity gradient. This radial velocity gradient can suppress the micro-meridional circulation until the driving force generated by the temperature difference balances the reaction generated by the radial velocity gradient. Ignoring the effects of viscosity, we have:
\begin{equation}
\oint \vec{F} \cdot d\vec{s} = 0
\end{equation}
Taking the streamline on the boundary as the research object, when there is a difference in latitude temperature, the gravity difference received by the region with a latitude difference of $d\varphi$ is:
\begin{equation}
F_g = gd\rho
\end{equation}
where
\begin{equation}
d\rho = -\rho \frac{dT}{T} = -\rho \frac{\partial T}{TR\partial \varphi} Rd\varphi
\end{equation}
The gravity difference can drive the micro-meridional circulation, and the direction of the Coriolis force generated by the micro-meridional circulation is perpendicular to the meridional plane. This can lead to changes in the radial differential rotation. The direction of the Coriolis force generated by this radial differential rotation is on the meridional plane and suppresses the circulation until it balances with the gravity difference. After balancing, the circulation velocity depends on the dissipation of the radial differential rotation by viscosity. When the viscosity is small, the circulation velocity is also small, but the meridional circulation driven by the gravity difference can maintain a larger value. This model is very similar to the drift of charged particles in a magnetic field under the influence of an electric field.

There are differences in the horizontal Coriolis force received by the fluid parcels on the upper and lower boundaries separated by $dR$, as well as differences in the Coriolis force on the high and low latitude boundaries. Their integrals on the loop are:
\begin{equation}
{\oint{\overset{\rightarrow}{F_{co}}d\overset{\rightarrow}{s}}} = 2\rho\Omega\left( {\frac{\partial{\delta v}_{\theta}}{\partial R}dRsin\varphi Rd\varphi + \frac{\partial{\delta v}_{\theta}}{R\partial\varphi}Rd\varphi cos\varphi dR} \right)
\end{equation}
After simplification, we obtain:
\begin{equation}
2\Omega\left(\frac{\partial \delta v_\theta}{\partial R} \sin\varphi + \frac{\partial \delta v_\theta}{R\partial \varphi} \cos\varphi\right) = -g \frac{\partial T}{TR\partial \varphi}
\end{equation}
This is the relationship (2) in the feedback chain. It can be seen that the latitudinal temperature gradient drives the micro-meridional circulation, causing additional radial differential rotation, which consumes energy.

The micro-meridional circulation driven by the latitudinal temperature gradient and the meridional circulation driven by the radial differential rotation are relatively independent, and the angular momentum transported by the micro-meridional circulation can be ignored.

The relationship between $\delta v_\theta$ and $\Omega$ or $\partial\Omega/\partial R$ in relationship (4) can be handled differently based on different assumptions. Generally, it can be assumed that the rotational speed difference caused by $\delta v_\theta$ is much smaller than the solar rotation speed $\Omega_0$. The rotational speed perpendicular to the rotation direction only causes a deviation in the rotation direction, and it is only necessary to project the rotational speed onto the rotation direction for calculation:
\begin{equation}
\Omega \approx \Omega_0 + \frac{\partial \delta v_\theta}{\partial R} \sin\varphi + \frac{\partial \delta v_\theta}{R\partial \varphi} \cos\varphi
\end{equation}
Differentiating, we obtain:
\begin{equation}
\frac{\partial \Omega}{\partial R} = \frac{\partial^2 \delta v_\theta}{\partial R^2} \sin\varphi + \frac{\partial^2 \delta v_\theta}{R\partial \varphi \partial R} \cos\varphi
\end{equation}
This is the relationship (4) in the feedback chain.

The critical size can be considered as the characteristic size of the fluid parcel. The influence of rotation speed and rotation speed gradient on the critical size of the fluid parcel in relationship (1) is given by the previous context:
\begin{equation}
l_{ad}^2 = \frac{-T\left(\frac{\partial T}{T\partial R} - (\gamma - 1) \frac{\partial\rho}{\rho\partial R}\right)}{\frac{2k_\Omega M}{R_m} \Omega^2 \left(\frac{\partial\Omega}{\Omega\partial R} - \frac{2}{3} \frac{\partial\rho}{\rho\partial R}\right)}
\end{equation}
By combining the four relationships, we can attempt to calculate the latitudinal temperature distribution.

Construct a thin spherical shell model and assume that all values of the model are within the linear range.

Relation (1) is differentiated with respect to $R\partial\varphi$ to obtain the rotational speed in the latitudinal direction, the rotational speed gradient, and the effect on the size of granules. In this relationship, the influence of changes in the temperature gradient is relatively small and can be ignored.
\begin{equation}
2\frac{\partial\Omega}{\Omega R\partial\varphi} + 2\frac{\partial l}{lR\partial\varphi} + \frac{\partial^{2}\Omega}{\left( {\frac{\partial\Omega}{\Omega\partial R} - \frac{2}{3}\frac{\partial\rho}{\rho\partial R}} \right)\Omega R\partial\varphi\partial R} \approx 0
\end{equation}
Relation (2) is differentiated with respect to $R\partial\varphi$ to obtain the temperature gradient changes caused by variations in granule size in the latitudinal direction.
\begin{equation}
\frac{\partial l}{lR\partial\varphi} + \frac{3}{2}\frac{\partial^{2}T}{\left( {\frac{\partial T}{T\partial R} - \left( {\gamma - 1} \right)\frac{\partial\rho}{\rho\partial R}} \right)TR\partial\varphi\partial R} = 0
\end{equation}
Combining relations (3) and (4), we have
\begin{equation}
\frac{\partial\Omega}{\partial R} = - \frac{g}{2\Omega sin\varphi}\frac{\partial^{2}T}{TR\partial\varphi\partial R}
\end{equation}
Incorporating the differentiation of relation (2), we obtain
\begin{equation}
\frac{\partial l}{lR\partial\varphi} = \frac{3\Omega}{g\left( {\frac{\partial T}{T\partial R} - \left( {\gamma - 1} \right)\frac{\partial\rho}{\rho\partial R}} \right)}\frac{\partial\Omega}{\partial R}
\end{equation}
Incorporating the differentiation of relation (1), we have
\begin{equation}
\begin{split}  
&2\frac{\partial\Omega}{\Omega R\partial\varphi} + \frac{6\Omega}{g\left( {\frac{\partial T}{T\partial R} - \left( {\gamma - 1} \right)\frac{\partial\rho}{\rho\partial R}} \right)}\frac{\partial\Omega}{\partial R} + \\
&\frac{\partial^{2}\Omega}{\left( {\frac{\partial\Omega}{\Omega\partial R} - \frac{2}{3}\frac{\partial\rho}{\rho\partial R}} \right)\Omega R\partial\varphi\partial R} \approx 0
\end{split}  
\end{equation}
The above equation depicts the feedback chain that results in the latitudinal temperature gradient approaching zero. The first term is determined by latitudinal differential rotation and is the origin of the problem. The second term is the rotational speed gradient determined by the latitudinal temperature gradient, and the third term is the total differential of the rotational speed, which is the residual effect of the second term.

Observational results show that, at least at the top of the solar convection zone, the latitudinal temperature gradient is close to 0. This means that the second term in the equation is only a bridge connecting the first and third terms. The area dominated by a single feedback chain is small, resulting in higher-order differentials being much larger than lower-order differentials. This only needs to prove that the feedback structure is sensitive to perturbations, and stronger or larger feedback structures will split into smaller and weaker structures when subjected to minor perturbations. This type of feedback mechanism can be proved to make the latitudinal temperature gradient tend to zero.

Re-examining relation (3), it can be found that this is only the simplest final stable state of the feedback structure, and this state requires viscosity to participate in dissipation. If the feedback structure extends infinitely in the $\theta$ direction, the feedback structure will be in an inertial oscillation state, and the oscillation period is determined by the rotation period and the shape of the feedback structure. However, during the oscillation process, the $v_\theta$ in different regions of the feedback structure is not the same, which will cause it to be torn apart quickly. This spontaneous disintegration will make the higher-order differentials in the feedback relationship significantly larger than the lower-order differentials, making the derived relationship consistent with observations.

The feedback structure has a tendency to disintegrate spontaneously, which makes it tend to have a smaller diameter. However, when the diameter of the feedback structure is close to the order of magnitude of the free path of fluid parcels, a large proportion of fluid parcels spanning different feedback structures will cause the feedback of the feedback structure to become less sensitive, and weaker feedback structures can be easily destroyed by stronger feedback structures. Therefore, the lower limit of the diameter of the feedback structure is related to the average free path of granules. The spontaneous disintegration of the feedback structure and the weakening of smaller feedback structures lead to a very concentrated size distribution of the feedback structure.

Let's briefly calculate the existence time of the feedback structure. The inertial oscillation period is half of the solar rotation period, which is $T_0/2$. The time for the structure to reach maximum speed from zero speed is $T_0/8\pi$, which can be used as the characteristic existence time of the feedback structure. The solar rotation period is approximately 27 days, so the existence time of the feedback structure is about 50 hours. This magnitude is very close to the existence time of the solar supergranulation structure. The gap between them may have the following possibilities: 1. The early stage of evolution is not obvious in observations, or it has not yet emerged to the surface of the convection zone. 2. Before the feedback structure reaches its extreme speed, the range of force application has changed, causing it to tear ahead of time.

In addition to the existence time, some characteristics of the feedback structure are also similar to the observed characteristics of supergranules. The size of granules in supergranules shows regular differences, which is a temperature regulation mechanism produced by the rotational speed gradient. This was previously thought to be caused by magnetic fields. The supergranules are dominated by horizontal motion and have significant differences from the convection structure, which corresponds to the small-scale but relatively high-speed meridional circulation and differential rotation in the feedback structure. From an energy perspective, this model also explains where the energy of the latitudinal temperature difference goes and where the energy of the supergranules comes from.

Supergranules can push weaker magnetic fields to their edges, but if the magnetic field is too strong, the feedback chain will be broken, causing temperature anomalies in that region and leading to abnormal feedback in nearby regions. This may be one of the mechanisms for the formation of solar activity.

%
%
%
%
%
%
%
%
%
%
%

\section{Conclusion} 

The rotational energy of a fluid parcel changes during isotropic expansion or compression which similar to temperature. When studying the effect of rotation on thermal convection, rotational energy can be described as rotational equivalent temperature.

The gradient of rotational equivalent temperature can influence thermal convection. Since rotational equivalent temperature is affected by the size $l$ of the fluid parcel, the convection criterion is related to $l$. For isotropic deforming fluid parcels, those smaller than a critical size $l_{ad}$ are in a convective state, while those larger than the critical size transition into oscillatory motion.

We have solved the equation of motion for a spherically symmetric deforming fluid parcel, neglecting viscosity. Fluid parcels larger than the critical size exhibit excited oscillatory solutions.

The deformation caused by turbulent irregular motion makes the equilibrium position of vibrating fluid parcels with $l>l_{ad}$ jump to a new place, generating motion. Thermal convection can also exist within them, which makes their photometric boundaries rough and their fractal dimension significantly larger than that of small granules. They are not normal thermal convection, so the relationship between their brightness and size is different from that of small granules.

Inertial oscillation can affect the rotational speed of fluid parcels, change the equilibrium position of fluid parcels in vibration, and form observable motion structures. After analyzing the movement of the vibrating equilibrium position, we found that the structural scale of this movement is concentrated near a certain value. This structure may be mesogranules.

The rotation of the fluid parcel absorbs energy from thermal convection, causing its rotational speed to differ from the environment. We have demonstrated through a significantly simplified model how a fluid parcel with a different rotational speed from the environment can generate radial differential rotation. This model yields higher speeds in the middle and lower speeds at the top and bottom of the convection zone, consistent with observations.

The Coriolis force from radial differential rotation drives meridional circulation and transports angular momentum towards the equator, forming latitudinal differential rotation. Based on the model of radial differential rotation, we have calculated the formation of latitudinal differential rotation using a simplified model. Its trend is close to the observed results.

The distribution of rotation rates, granule sizes, temperature distributions, and micro-meridional circulations constitute a complete feedback chain. This chain drives the latitudinal temperature gradient to nearly zero and the granule sizes to be nearly uniform.  The life time and velocity characteristics of the feedback structure are similar to those of supergranules. The disruption of the feedback chain by magnetic fields may be one of the mechanisms for the formation of solar activity.

\section*{acknowledgments}
The data underlying this article are available in the article and in its online supplementary material.
This paper was translated by AI Wenxinyiyan.

\bibliography{reference}{}

\begin{thebibliography}{}
\expandafter\ifx\csname natexlab\endcsname\relax\def\natexlab#1{#1}\fi
\providecommand{\url}[1]{\href{#1}{#1}}
\providecommand{\dodoi}[1]{doi:~\href{http://doi.org/#1}{\nolinkurl{#1}}}
\providecommand{\doeprint}[1]{\href{http://ascl.net/#1}{\nolinkurl{http://ascl.net/#1}}}
\providecommand{\doarXiv}[1]{\href{https://arxiv.org/abs/#1}{\nolinkurl{https://arxiv.org/abs/#1}}}

\bibitem[{{Beckers}(1968)}]{1968SoPh....5..309B}
{Beckers}, J.~M. 1968, \solphys, 5, 309, \dodoi{10.1007/BF00147143}

\bibitem[{{Chen} \& {Wu}(2022{\natexlab{a}})}]{2022arXiv220711990C}
{Chen}, H., \& {Wu}, R. 2022{\natexlab{a}}, arXiv e-prints, arXiv:2207.11990,
  \dodoi{10.48550/arXiv.2207.11990}

\bibitem[{{Chen} \& {Wu}(2022{\natexlab{b}})}]{2022arXiv221108113C}
---. 2022{\natexlab{b}}, arXiv e-prints, arXiv:2211.08113,
  \dodoi{10.48550/arXiv.2211.08113}

\bibitem[{{Deubner}(1989)}]{1989A&A...216..259D}
{Deubner}, F.-L. 1989, \aap, 216, 259

\bibitem[{{Frazier}(1970)}]{1970SoPh...14...89F}
{Frazier}, E.~N. 1970, \solphys, 14, 89, \dodoi{10.1007/BF00240163}

\bibitem[{{Gilman}(1974)}]{1974ARA&A..12...47G}
{Gilman}, P.~A. 1974, \araa, 12, 47,
  \dodoi{10.1146/annurev.aa.12.090174.000403}

\bibitem[{{Hart}(1954)}]{1954MNRAS.114...17H}
{Hart}, A.~B. 1954, \mnras, 114, 17, \dodoi{10.1093/mnras/114.1.17}

\bibitem[{{Hirzberger} {et~al.}(1997){Hirzberger}, {V{\'a}zquez}, {Bonet},
  {Hanslmeier}, \& {Sobotka}}]{1997ApJ...480..406H}
{Hirzberger}, J., {V{\'a}zquez}, M., {Bonet}, J.~A., {Hanslmeier}, A., \&
  {Sobotka}, M. 1997, \apj, 480, 406, \dodoi{10.1086/303951}

\bibitem[{{Howard} {et~al.}(1984){Howard}, {Gilman}, \&
  {Gilman}}]{1984ApJ...283..373H}
{Howard}, R., {Gilman}, P.~I., \& {Gilman}, P.~A. 1984, \apj, 283, 373,
  \dodoi{10.1086/162315}

\bibitem[{{Hoyt} \& {Schatten}(1993)}]{1993JGR....9818895H}
{Hoyt}, D.~V., \& {Schatten}, K.~H. 1993, \jgr, 98, 18895,
  \dodoi{10.1029/93JA01944}

\bibitem[{{Kuhn} {et~al.}(1988){Kuhn}, {Libbrecht}, \&
  {Dicke}}]{1988Sci...242..908K}
{Kuhn}, J.~R., {Libbrecht}, K.~G., \& {Dicke}, R.~H. 1988, Science, 242, 908,
  \dodoi{10.1126/science.242.4880.908}

\bibitem[{{Leighton} {et~al.}(1962){Leighton}, {Noyes}, \&
  {Simon}}]{1962ApJ...135..474L}
{Leighton}, R.~B., {Noyes}, R.~W., \& {Simon}, G.~W. 1962, \apj, 135, 474,
  \dodoi{10.1086/147285}

\bibitem[{{Maunder} \& {Maunder}(1905)}]{1905MNRAS..65..813M}
{Maunder}, E.~W., \& {Maunder}, A.~S.~D. 1905, \mnras, 65, 813,
  \dodoi{10.1093/mnras/65.8.813}

\bibitem[{{Newton} \& {Nunn}(1951)}]{1951MNRAS.111..413N}
{Newton}, H.~W., \& {Nunn}, M.~L. 1951, \mnras, 111, 413,
  \dodoi{10.1093/mnras/111.4.413}

\bibitem[{{November} {et~al.}(1981){November}, {Toomre}, {Gebbie}, \&
  {Simon}}]{1981ApJ...245L.123N}
{November}, L.~J., {Toomre}, J., {Gebbie}, K.~B., \& {Simon}, G.~W. 1981,
  \apjl, 245, L123, \dodoi{10.1086/183539}

\bibitem[{{Patern{\`o}}(2010)}]{2010Ap&SS.328..269P}
{Patern{\`o}}, L. 2010, \apss, 328, 269, \dodoi{10.1007/s10509-009-0218-0}

\bibitem[{{Rast}(2003)}]{2003ApJ...597.1200R}
{Rast}, M.~P. 2003, \apj, 597, 1200, \dodoi{10.1086/381221}

\bibitem[{{Richardson} \& {Schwarzschild}(1950)}]{1950ApJ...111..351R}
{Richardson}, R.~S., \& {Schwarzschild}, M. 1950, \apj, 111, 351,
  \dodoi{10.1086/145269}

\bibitem[{{Roudier} \& {Muller}(1986)}]{1986SoPh..107...11R}
{Roudier}, T., \& {Muller}, R. 1986, \solphys, 107, 11,
  \dodoi{10.1007/BF00155337}

\bibitem[{{Scherrer} {et~al.}(1995){Scherrer}, {Bogart}, {Bush}, {Hoeksema},
  {Kosovichev}, {Schou}, {Rosenberg}, {Springer}, {Tarbell}, {Title},
  {Wolfson}, {Zayer}, \& {MDI Engineering Team}}]{1995SoPh..162..129S}
{Scherrer}, P.~H., {Bogart}, R.~S., {Bush}, R.~I., {et~al.} 1995, \solphys,
  162, 129, \dodoi{10.1007/BF00733429}

\bibitem[{{Snodgrass}(1984)}]{1984SoPh...94...13S}
{Snodgrass}, H.~B. 1984, \solphys, 94, 13, \dodoi{10.1007/BF00154804}

\bibitem[{{Straus} {et~al.}(1992){Straus}, {Deubner}, \&
  {Fleck}}]{1992A&A...256..652S}
{Straus}, T., {Deubner}, F.~L., \& {Fleck}, B. 1992, \aap, 256, 652

\bibitem[{{Unno} {et~al.}(1989){Unno}, {Osaki}, {Ando}, {Saio}, \&
  {Shibahashi}}]{1989nos..book.....U}
{Unno}, W., {Osaki}, Y., {Ando}, H., {Saio}, H., \& {Shibahashi}, H. 1989,
  {Nonradial oscillations of stars}

\bibitem[{{Uns{\"o}ld}(1930)}]{1930ZA......1..138U}
{Uns{\"o}ld}, A. 1930, \zap, 1, 138

\bibitem[{{Weiss}(1965)}]{1965Obs....85...37W}
{Weiss}, N.~O. 1965, The Observatory, 85, 37

\bibitem[{{Willson} \& {Hudson}(1988)}]{1988Natur.332..810W}
{Willson}, R.~C., \& {Hudson}, H.~S. 1988, \nat, 332, 810,
  \dodoi{10.1038/332810a0}

\end{thebibliography}
\bibliographystyle{aasjournal}

\end{CJK*}
\end{document}